\documentclass[nologo,11pt,a4paper]{ETHpaper}     
\usepackage[pdftex]{graphicx}
\usepackage{subfigure}
\usepackage{amsmath}
\usepackage{multirow}
\usepackage{color, colortbl}
\usepackage{afterpage}
\usepackage{url} 
\definecolor{Gray}{rgb}{0.9,0.9,0.9}

\begin{document}   

\title{The Rise and Fall of a Central Contributor: Dynamics of Social Organization and Performance in the \textsc{Gentoo} Community}
\titlealternative{The Rise and Fall of a Central Contributor:\\Dynamics of Social Organization and Performance in the \textsc{Gentoo} Community}
\author{Marcelo Serrano Zanetti, Ingo Scholtes,\\ Claudio Juan Tessone and Frank Schweitzer}
\authoralternative{Marcelo Serrano Zanetti, Ingo Scholtes, Claudio Juan Tessone and Frank Schweitzer}
\address{Chair of Systems Design, ETH Zurich, Switzerland\\
  \url{www.sg.ethz.ch}} 
\reference{to appear in the proceedings of the 6th International Workshop on Cooperative and Human Aspects of Software Engineering (CHASE 2013) - ICSE 2013 Workshop}
\www{\url{http://www.sg.ethz.ch}} 
\makeframing
\maketitle 

\begin{abstract}
Social organization and division of labor crucially influence the performance of collaborative software engineering efforts.
In this paper, we provide a quantitative analysis of the relation between  social organization and performance in \textsc{Gentoo}, an Open Source community developing a \textsc{Linux} distribution.
We study the structure and dynamics of collaborations as recorded in the project's bug tracking system over a period of ten years.
We identify a period of increasing centralization after which most interactions in the community were mediated by a single central contributor.
In this period of maximum centralization, the central contributor unexpectedly left the project, thus posing a significant challenge for the community.
We quantify how the rise, the activity as well as the subsequent sudden dropout of this central contributor affected both the social organization and the bug handling performance of the \textsc{Gentoo} community.
We analyze social organization from the perspective of network theory and augment our quantitative findings by interviews with prominent members of the \textsc{Gentoo} community which shared their personal insights.
\end{abstract}

\section{Introduction}
\label{sec:introduction}

An important prerequisite for the success of Open Source Software (OSS) projects is the ability to build a sufficiently large and stable community of users and contributors.
While actual source code is typically contributed by a rather small and stable set of core developers,
the wider - and possibly more diverse - community plays an important role  in processes related to software quality management.
Here, most OSS projects rely on a large number of part-time \emph{contributors} who report bugs, triage pending bug reports or provide support and solutions for issues reported by others.
This community effort in handling bug reports does not only unburden developers; it also significantly increases software quality, thus bearing the potential to attract more users.
Furthermore, in \cite{Zhou2012} it was argued that - despite their volatility - bug handling communities are a common entry point for long-term contributors who - after getting insight into a project's organizational and technical structures - may eventually become members of the core developer team.
As such, the structure and dynamics of bug handling communities is of particular importance for the success of OSS projects.
In order to ensure a timely response to bug reports, the management of the project has to find efficient organizational structures and a reasonable division of labor, despite the fact that these communities are typically  highly heterogeneous in terms of dedication and skills.

In this paper, we present a quantitative analysis of the structure and dynamics of the bug handling community of \textsc{Gentoo}, an OSS project developing a \textsc{Linux} distribution.
Our study is based on a data set covering more than $150,000$ collaboration events recorded by the project's \textsc{Bugzilla} installation over a period of more than ten years.
The contributions of our study are as follows:

\begin{itemize}
  \item We study collaboration structures of the \textsc{Gentoo} bug handling community by applying quantitative measures that capture cohesion, centralization, clustering and communication efficiency.
      Our analysis reveals a period of increasing centralization and decreasing cohesion that resulted in a situation where most interactions in the community were mediated by a single \emph{central contributor}.
  \item In the period of maximum centralization the central contributor unexpectedly left the project.
  We analyze the implications for the project's social organization, which include a temporary loss of cohesion as well as subsequent efforts to reorganize the community.
  \item We complement our study by an analysis of the community's performance in terms of bug handling efficiency and response time.
Our findings suggest that the performance improved during the active period of the \emph{central contributor}, while her retirement had a lasting negative effect on bug handling efficiency and response time.
  \item We substantiate our quantitative findings by personal insights into the social dynamics of the \textsc{Gentoo} community provided by three long-term contributors.
      These insights support our findings and highlight potential applications of our quantitative measures in the monitoring of collaboration structures in OSS projects.
\end{itemize}

The remainder of this paper is organized as follows.
In section \ref{sec:related} we summarize relevant related work studying the structure of OSS communities and its impact on performance.
In section \ref{sec:methods} we introduce data collection and network analysis methods that form the basis of our case study.
In section \ref{sec:results} we present quantitative results on the evolution of the social organization, as well as bug handling performance in the \textsc{Gentoo} community.
We further interpret our findings, align them with personal insights shared by prominent community members and discuss threats to validity.
Finally, in section \ref{sec:conclusion} we summarize our contributions and highlight future research on the application of network-based analysis methods in the management of software development communities.

\section{Studies of Social Organization in Collaborative Software Engineering}
\label{sec:related}
%
The question how the structure and dynamics of social organization influences the performance and success of collaborative software development efforts has been studied by researchers from different fields using a variety of methods. 
Due to the availability of data, many of these studies address OSS communities, which consist of \emph{users}, \emph{developers} and other \emph{contributors}, who contribute to the project in terms of documentation, maintenance of web sites or the submission and handling of bug reports.
Members of such communities typically need to self-organize in a way that guarantees information flow as well as a coordinated allocation of tasks and responsibilities.
The processes and structures of this self-organization process have been studied in a number of works.
 
Since it plays a central role in software quality assurance, \emph{bug handling communities} have been the subject of many studies.
Compared to the development of source code, in \cite{Mockus2002} it was found that the bug handling process is based on the contributions of a much wider community.
In a recent work presented in \cite{Zhou2012}, this community has further been shown to be an important entry point for long-term contributors and developers.
As an important finding, lack of attention paid to bug reporters and fast negative feedback by the community decreases the likelihood for such users to contribute to the project for a long period.
This is partly in line with arguments about the negative impact of a too strict duplicate bug policy in bug handling communities put forth in \cite{Bettenburg2008a}.
  
The collaboration structures emerging in bug handling communities can be extracted by different means.
Communication topologies of the bug handling communities of OSS projects hosted on \textsc{SourceForge} have been analyzed in \cite{Crowston2005}.
Here it was shown that large projects - measured in terms of the number of contributors - tend to have lower degrees of centralization in communication.
The authors further call for a detailed longitudinal analysis of changes in the social organization of OSS projects during periods of growth.
Our work complements this study in the sense that we a) analyze the dynamics of centralization during a phase of growth in the \textsc{Gentoo} community and b) show the impact of increasing centralization on community performance and cohesion. 

In addition to studies at the level of the community, the relationship between the network position of contributors and their individual success (like e.g. the number of bug reports leading to bug fixes) has been studied in \cite{Ehrlich2012}.
The authors find that both the centrality of contributors, as well as their embedding in cohesive clusters of communication has beneficial effects on the bug fixing performance.
A similar finding has been presented in \cite{Zanetti2013}, which studies the impact of social aspects on individual performance in bug handling communities.
Our paper complements this work in the sense that we study network-wide measures of communication efficiency and centralization, their dynamics during community growth, as well as their relation to the bug handling performance of the community.
We further highlight potential risks associated with the presence of central contributors in situations when these contributors leave the community unexpectedly.

The relationship between communication structures and success at the level of teams was studied in \cite{Wolf2009}.
Here it was shown that positive team performance is related to communication structures that facilitate information dissemination.
However, no clear relation between differences in the coordination practices and the project success could be identified.
In \cite{Zanetti2012}, the dynamics of collaboration structures of $14$ OSS communities has been studied.
Similarly, in \cite{Crowston2004}, co-ordination practices of the bug handling process have been studied for four OSS communities.
The authors found that contributions are not distributed equally and that the community is organized in a core-periphery structure.  
Unequal division of labor and an increasing degree of centralization are compatible with findings about the rise of a leader are presented in \cite{giuri2008}.
Here, a leader is defined as a contributor who consistently provides high quality contributions, co-ordinates efforts \cite{scozzi2008} and around whom the community is centered \cite{evans2005}.
Usually, leadership in OSS projects is shared between several contributors.
The analysis performed in \cite{Sadowski2008} shows that overdependence on a leader results in an unstable situation where the project may accelerate - initially - its development, but which may end up saturating the leader.

The present paper extends these previous works in the following way. First, we study the dynamics of a more comprehensive set of network measures that can be interpreted in terms of \emph{cohesion}, \emph{centralization} and \emph{communication efficiency}.
We particularly study how the social organization of the \textsc{Gentoo} community evolves during an initial phase of growth and a subsequent phase of increasing centralization that is due to the presence of a central contributor. 
We then relate our results with proxies for community performance and study how both performance and social organization are impacted by the loss of a central contributor. 
Finally, we interpret and substantiate our findings by means of insights from actual contributors to the \textsc{Gentoo} community.

\section{Methodology}
\label{sec:methods}


In our study of the dynamics of social organization in the bug handling community of \textsc{Gentoo}, we use the project's installation of the \textsc{Bugzilla} bug tracker as data source.
We first describe our process of retrieving data and extracting evolving collaboration networks.
We then introduce the quantitative measures applied in our analysis of collaboration networks and briefly comment on their interpretation in the context of OSS projects.
Furthermore, we summarize how we selected three community members in order to substantiate our findings by means of personal insights from former and active contributors to the \textsc{Gentoo} project.

\subsection{Data Collection}

In January 2002, the \textsc{Gentoo} community started to use the \textsc{Bugzilla} bug tracking system.
The full history of all bug reports submitted since then are recorded in the database of the project's \textsc{Bugzilla} installation.
Data available for each of these bug reports include the history of all updates to any field along with time stamps and the ID of the user who applied the update.
In the context of our analysis, we particular extract the ID of the user who initially submitted a bug report, as well as the time of the submission and the status of a bug report, like e.g. \emph{unconfirmed}, \emph{pending}, \emph{reproduced} or \emph{resolved}.
For those bugs whose final status was set to \emph{resolved}, we additionally collected the \emph{resolution} field of the report, which can take one of the values \emph{fixed}, \emph{duplicate}, \emph{invalid}, \emph{needinfo} or \emph{wontfix}.
An entry \emph{fixed} refers to those bugs for which the community eventually provided a fix.
Bug reports whose \emph{resolution} field was set to \emph{duplicate} were identified to be duplicates of an existing bug report that refers to the same issue.
Bugs with the final resolution \emph{invalid} are those that do not refer to actual software issues, instead referring for instance to a misunderstanding on the user's side.
If a bug report is incomplete in the sense that it lacks important information that would allow to reproduce or fix the underlying issue and if the reporting user fails to provide the necessary information within a certain time, the \emph{resolution} field of a bug report is set to \emph{needinfo}.
Finally, the resolution of those bug reports that are valid and complete, but that nevertheless cannot be fixed either due to a lack of resources or the fact that the issue is due to a external dependency are marked as \emph{wontfix}.
The fact that all changes to the \emph{resolution} field of a bug report as well as the submission of the bug report itself are associated with a precise time stamp, further allows us to compute the number of bugs that were submitted or resolved with a given status within a given period of time.
In addition to all updates that relate to the resolution status of a bug, we also extracted the full history of the \emph{assignee} and the \emph{cc} fields of each bug report.
The \emph{assignee} field contains the ID of the user who was made responsible for providing a solution for a particular bug report, while the \emph{cc} field contains a list of user IDs that are being notified about any future updates on a particular bug.

All of the data were collected via the public \emph{API} of the \textsc{Gentoo} project's \textsc{Bugzilla} installation.
In total, we retrieved data on $140,216$ bug reports and $661,783$ change events recorded between January 1st 2002 and April 26th 2012.
Some statistics of the data set, including the fraction of resolved bugs falling in each of the aforementioned resolution categories are shown in Table \ref{tab:Statistics}.
    
\subsection{Network Construction}

A core aspect of our study is the quantitative analysis of the collaboration structure of the \textsc{Gentoo} community during particular periods of time.
Even though our data set contains the full record of updates to bug reports, for the construction of collaboration networks, we limit our study to those update events that unambiguously capture dyadic \emph{social} interactions between two contributors.
In particular, for each addition of a user ID to the \emph{cc} and \emph{assignee} field of a bug report, we infer a dyadic interaction between the contributor performing the change and the ID of the user that was added to the field.
We further associate this interaction with the time stamp of the associated update of the bug report.
Focusing on updates to the \emph{cc} and \emph{assignee} fields of bug reports necessarily provides a limited perspective on the social organization of a community.
Nevertheless we decided to neglect additional data like e.g. the sequence of comments on bugs for which an inference of directed interaction networks is more difficult and error-prone.
We rather argue that the collaboration networks resulting from our construction procedure are nevertheless insightful.
The fact that a contributor $A$ adds contributor $B$ to the \emph{cc} field of a bug indicates that $A$ is aware of $B$ and that $A$ knows about the interests or competencies of $B$.
Furthermore, the fact that contributor $X$ adds $Y$ to the \emph{assignee} field of a bug report highlights that these contributors have different roles in the community, like e.g. $X$ identifying the cause of an issue and assigning it to $Y$.

\begin{figure}[ht]
\centering
\includegraphics[width=0.7\textwidth]{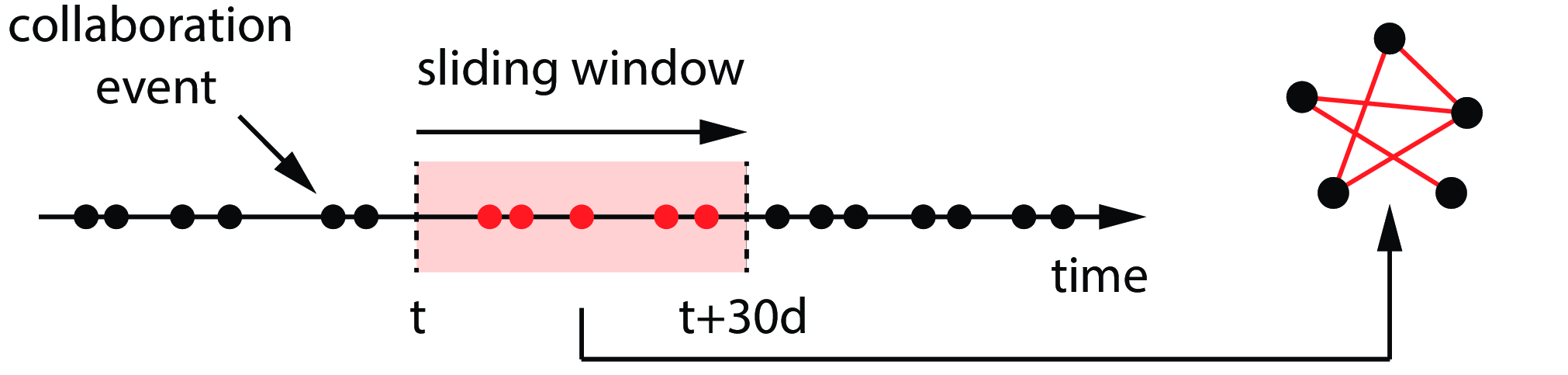}
\caption{Sliding window procedure for the construction of evolving networks.\label{fig:SlidingWindow}}
\end{figure}

Excluding those change events where contributors added themselves to the \emph{cc} or \emph{assignee} field, we infer more than $150,000$ directed interactions between different members of the \textsc{Gentoo} community.
The structure and dynamics of these interactions can be studied in terms of a \emph{collaboration network} in which nodes represent contributors and directed edges represent interactions between them.
A quantitative analysis of such network structures can reveal interesting insights into the community's organization.
Rather than aggregating all interactions occurring over a period of ten years, we further utilize the fact that all interactions inferred from our data set are \emph{time-stamped}.
In particular, we define a time window of $30$ days, filter out all interactions whose time stamps are outside this time window and construct a network from all remaining interactions (see an illustration of this procedure in Figure \ref{fig:SlidingWindow}).
Starting on the first day of the observation period, we then progressively slide the start date of this time window by one day increments.
This sliding window approach yields a sequence of $3,765$ networks, each of them representing the collaboration structures of the community within a $30$ day period starting at a particular day.
By analyzing this sequence of networks, we obtain a time series of network measures that capture the dynamics of social organization.
It is important to note that the collaboration networks obtained in the way described above are not necessarily connected, i.e. they may consist of different disconnected components.
In order to still provide a single measure that can be compared to previous and subsequent snapshots, we limit our analysis to the network's \emph{largest connected component} (LCC).
We additionally measure the size of the LCC and indicate its relative size in terms of the fraction of all nodes that are connected to the LCC.

\begin{table}[!t]
\centering
\caption{Basic statistics of the \textsc{Bugzilla} data set used for this study.\label{tab:Statistics}}
\begin{tabular}{l|r}
     Statistic & \textsc{Gentoo} \\
     \hline
     \rowcolor{Gray}                                         & 01/04/2002\\
     \rowcolor{Gray}    \multirow{-2}{*}{Observation period}& to 04/26/2012 \\
                        Bug reports                         & 140,216 \\
     \rowcolor{Gray}    Change events                       & 661,783 \\
                        Users                               & 36,555 \\
     \rowcolor{Gray}    Collaboration events                & 153,610 \\
                        Change events / Bug reports         &  4.72 \\
     \rowcolor{Gray}    Resolved (Resolved / Bug reports)   & 86,352 (0.61) \\
     \hline
     \rowcolor{Gray}    FIXED (FIXED / Resolved)            & 39,858 (0.46) \\
                        DUPLICATE (DUPLICATE / Resolved)    & 20,529 (0.24) \\
     \rowcolor{Gray}    INVALID (INVALID / Resolved)        & 14,923 (0.17) \\
                        WONTFIX (WONTFIX / Resolved)        &  7,959 (0.09) \\
     \rowcolor{Gray}    NEEDINFO (NEEDINFO / Resolved)      &  3,083 (0.04)\\
   \end{tabular}
\end{table}

\subsection{Network Measures}

In the following, we briefly introduce a number of quantitative, network-theoretic measures that we found to capture interesting aspects of the dynamics of social organization in the \textsc{Gentoo} community.
For the sake of brevity, we omit the formal definition of these measures and rather introduce the  motivation and interpretation in the context of our study.
For the formal definition of the measures mentioned in this section, we refer the interested reader to \cite{Wasserman1994,Newman2010}.
For our analysis we use their respective implementations in the network analysis package \textsc{igraph} \cite{Csardi2006}.

\paragraph{Closeness centralization} The first measure that we applied in our analysis is \emph{closeness centrality}.
The normalized closeness centrality of a node can be defined based on the inverse of the sum of the shortest path lengths to all other nodes in the network.
As such, it captures the centrality of a node in terms of how close it is to all other nodes in the network.
Based on the distribution of \emph{closeness centralities} of all nodes, one can furthermore define the so-called \emph{closeness centralization} of a network.
This network-wide measure captures the degree to which the topology is \emph{centralized}.
In a (maximally centralized) star network it takes a maximum value of $1$ while it is $0$ for networks in which all shortest paths between all pairs of nodes have the same length (like e.g. a fully connected topology).
In the context of our analysis, the closeness centralization of a collaboration network captures to what degree contributors have the same importance for indirect information exchange.
Precisely, in a network with maximum closeness centralization all collaborations are mediated by a single individual, while in networks with smaller closeness centralization community members have more equal roles.

\paragraph{Clustering coefficient} The \emph{clustering coefficient} of a network measures how closely community members interact with each other in the sense that interactions between users $A$ and $B$, as well as between $B$ and $C$ will also entail a direct interaction between the users $A$ and $C$.
This property of a network can be quantified at the level of nodes by computing the fraction of those pairs of a node's neighbors $u$ and $v$ that are connected by a direct link $(u,v)$.
By averaging the clustering coefficient scores of all nodes it is possible to measure the global clustering coefficient of a network.
In the context of our analysis, the (mean) clustering coefficient of a monthly collaboration network captures how \emph{cohesive} the community is in terms of contributors being embedded in collaborating clusters.
In other words, this measure captures to what extent two collaborators also collaborate with other collaborators of their peers.

\paragraph{Degree Assortativity} The \emph{degree assortativity} of a node measures an individual's preference to connect to peers that have a similar or different number of connections (degree).
Networks in which nodes are preferentially connected to nodes with similar degree are called assortative.
A positive degree assortativity indicates a positive correlation between the degrees of neighboring nodes.
Networks in which nodes are preferentially connected to nodes with different (i.e. smaller or higher) degree are called disassortative, in which case degree assortativity is negative.
Zero \emph{degree assortativity} means that there is no correlation between the degrees of connected nodes, i.e. nodes do not exhibit a preference for one or the other.
In our analysis, we use degree assortativity to capture the contributors' preference to collaborate with other contributors that are - from the perspective of the number of collaborations - of similar or different importance than themselves.

\paragraph{Algebraic Connectivity} An interesting aspect of network analysis is that the influence of a network topology on \emph{dynamical processes} like e.g. information flow, cascading failures or synchronization phenomena can be captured by means of so-called \emph{spectral} properties.
One important measure in this line is the so-called \emph{algebraic connectivity} of a network.
This scalar property particularly captures whether the topology contains \emph{small cuts}, i.e. whether all shortest paths between different parts of the network pass through a small number of edges.
The existence of such small cuts is known to hinder information spreading and synchronization \cite{graphtheory2001}.
At the same time, it can be seen as an indicator for robustness since it captures the effect of a failure of a small number of nodes and associated links.
The algebraic connectivity is defined as being the second smallest eigenvalue of the so-called Laplacian matrix, which is defined as the difference $D-A$ between a diagonal matrix $D$ in which the diagonal elements represent the degrees of nodes and the adjacency matrix $A$ of the network topology.
The algebraic connectivity of a network is greater than 0 \emph{iff} the network topology is connected, i.e. iff a path exists between all pairs of nodes.
This is a corollary to the fact that the number of times $0$ appears as an eigenvalue of the Laplacian matrix is equal to the number of the network's connected components.
In the context of this paper we use algebraic connectivity it to measure the communication efficiency and robustness of the the community's collaboration structure.

\subsection{Interviews with community members}
In order to substantiate our quantitative findings with insights into the community, we contacted a number of long-term contributors to the \textsc{Gentoo} bug handling community.
We received three very insightful and detailed replies, which contain many details and serve as an external validation for our quantitative findings.
We omit the real names of the contributors and refer to them as \emph{Alice}, \emph{Bob} and \emph{Chris} instead.
\emph{Alice} was the - by far - most central contributor to the \textsc{Gentoo} bug handling community in the period between October 2004 and March 2008.
She was effectively handling most of the bug reports, until she left the project suddenly in March 2008.
\emph{Bob} was involved in a community initiative to establish formal procedures regarding the submission and handling of bug reports that were - in part - necessitated by the departure of \emph{Alice}.
\emph{Chris} is another long-term contributor to the project, second only to \emph{Alice} in terms of cumulative contributions to the bug handling process.
In our questionnaire, we asked for personal insights regarding the following questions:
\begin{itemize}
  \item What was the impact of the central contributor \emph{Alice} on the involvement of other contributors and project performance?
  \item What were the reasons for \emph{Alice} leaving the project?
  \item What was the motivation for the establishment of formal procedures for the bug handling process?
  \item Was this initiative successful in terms of improving the performance of the community?
  \item What implications did the establishment of formal procedures have for the social organization of the community?
\end{itemize}

\section{Dynamics of Social Organization and Performance}
\label{sec:results}

In the following, we study the dynamics of \textsc{Gentoo}'s bug handling community during the time between 2002 and 2012.
Our methodology is based on a network-theoretic analysis of collaboration networks by means of the measures discussed in section \ref{sec:methods}.

Based on the activity of the central contributor \emph{Alice}, we divide the observation period into three periods $P1$, $P2$, $P3$.
In period $P1$, between January 2002 and October 27, 2004, \emph{Alice} was not yet active and the community was growing.
During the second period $P2$ starting on October 28 2004, \emph{Alice} gradually became the most central contributor.
She unexpectedly left the community after her last contribution on March 29 2008, which marks the start of the third period $P3$ in which $Alice$ was not active anymore.
In the following sections, we show how community cohesion, centralization and performance evolved in these three periods.

\subsection{Community Cohesion}
\label{sec:cohesion}  
\begin{figure}[!ht]
\centering
\subfigure[network size\label{fig:structure:size}]{
  \includegraphics[width=0.3\textwidth]{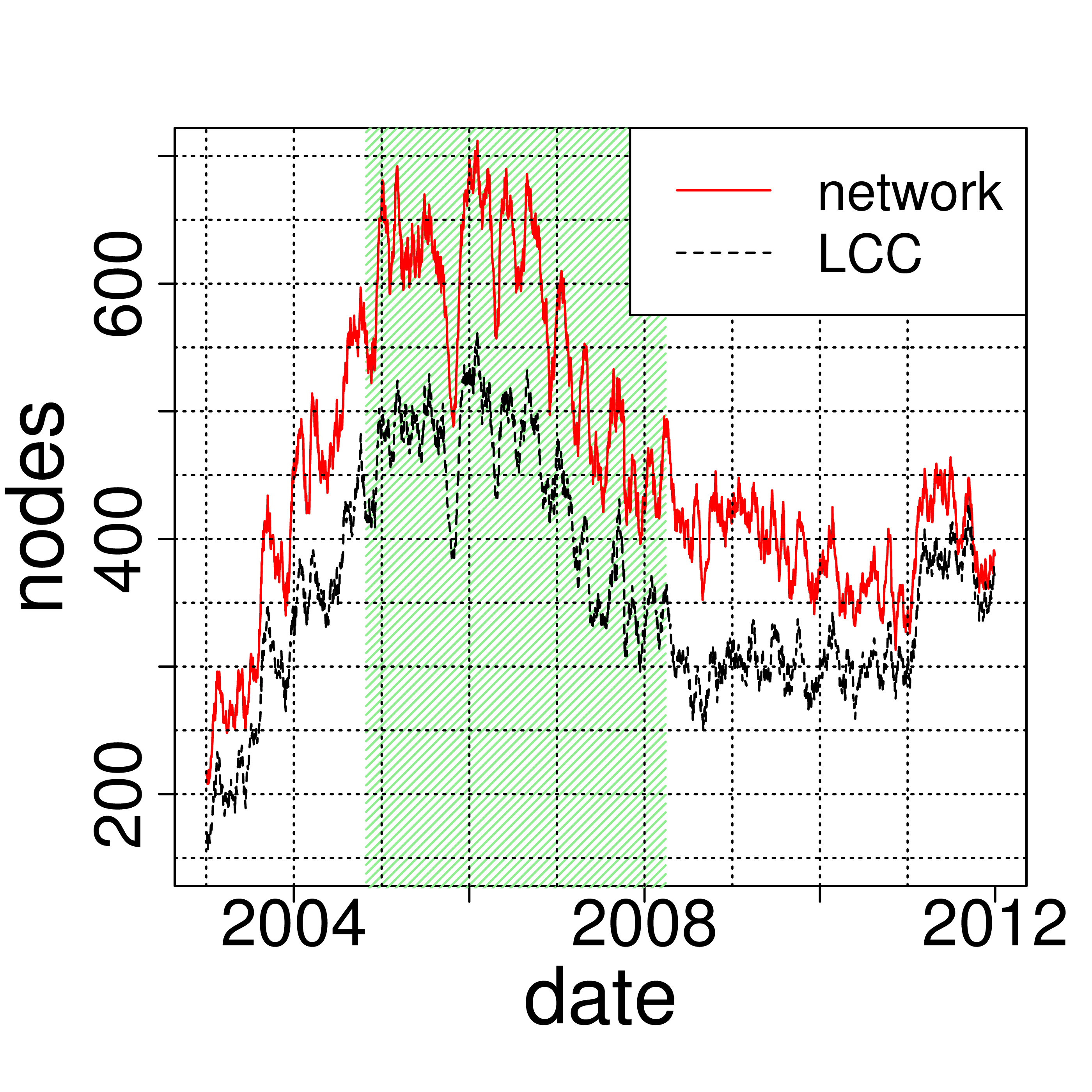}
}
\subfigure[relative size\label{fig:structure:lcc_fraction}]{
  \includegraphics[width=0.3\textwidth]{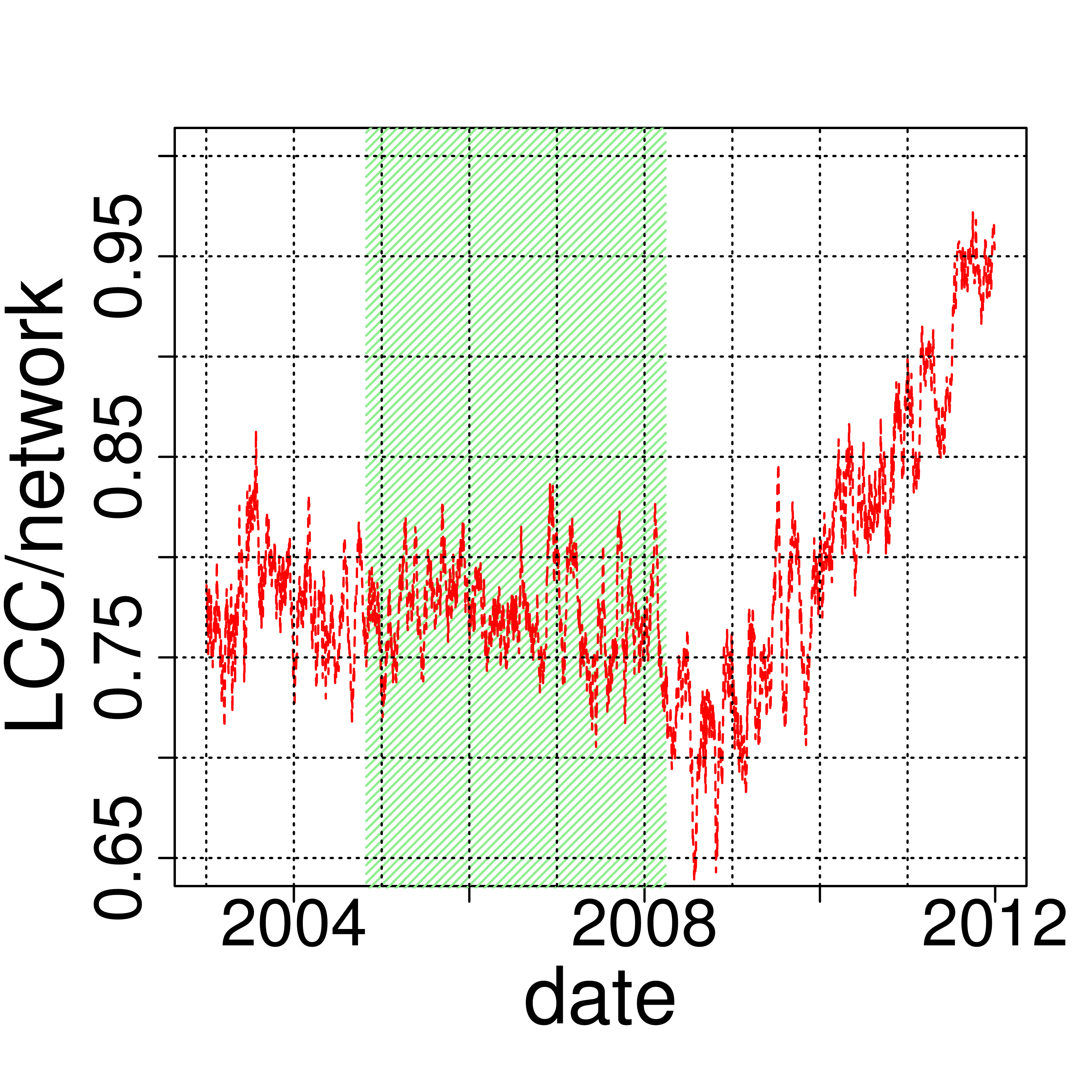}
}
\subfigure[relative size (without \emph{Alice})\label{fig:structure_wo_alice:lcc_fraction}]{
  \includegraphics[width=0.3\textwidth]{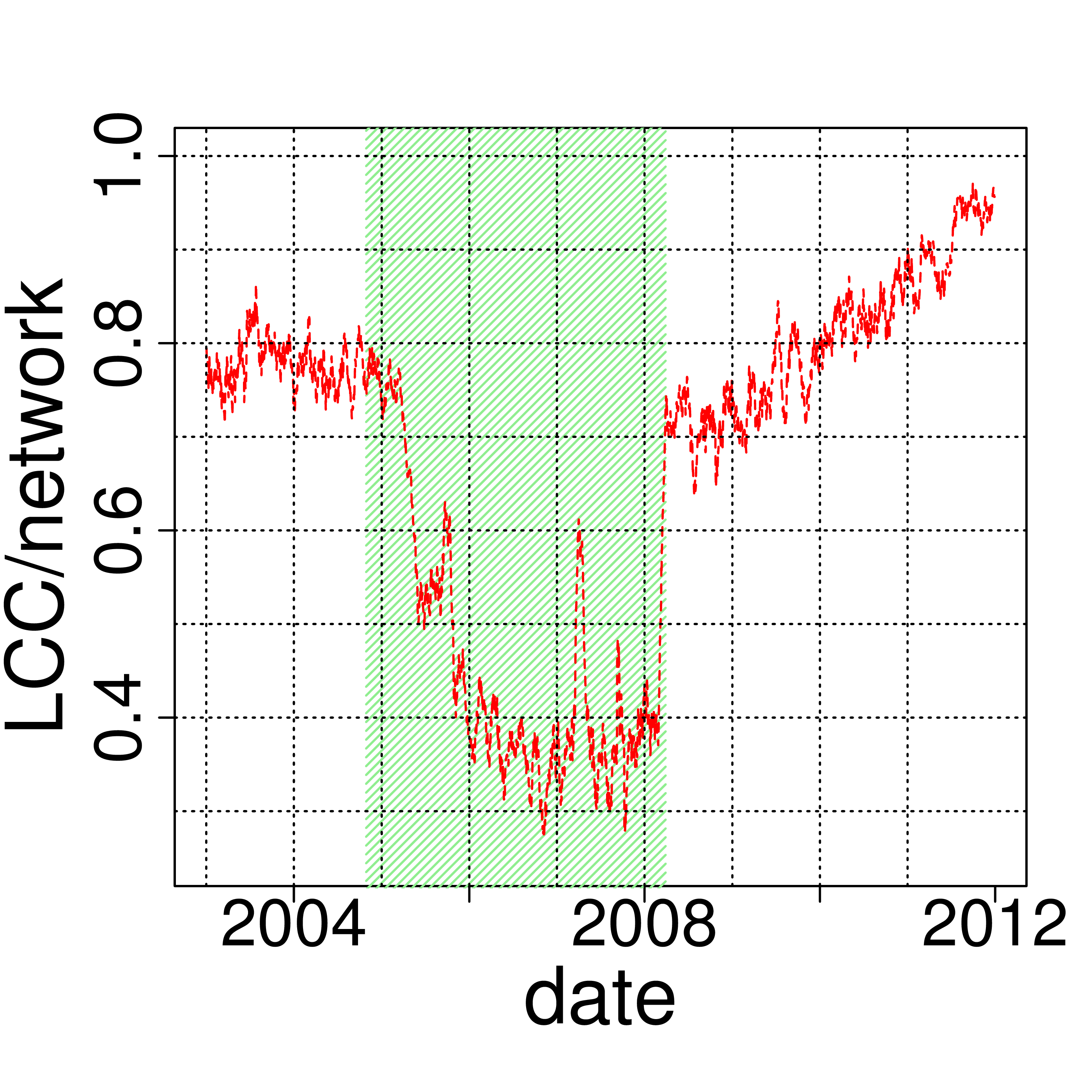}
}
\\
\subfigure[mean degree\label{fig:structure:md}]{
  \includegraphics[width=0.3\textwidth]{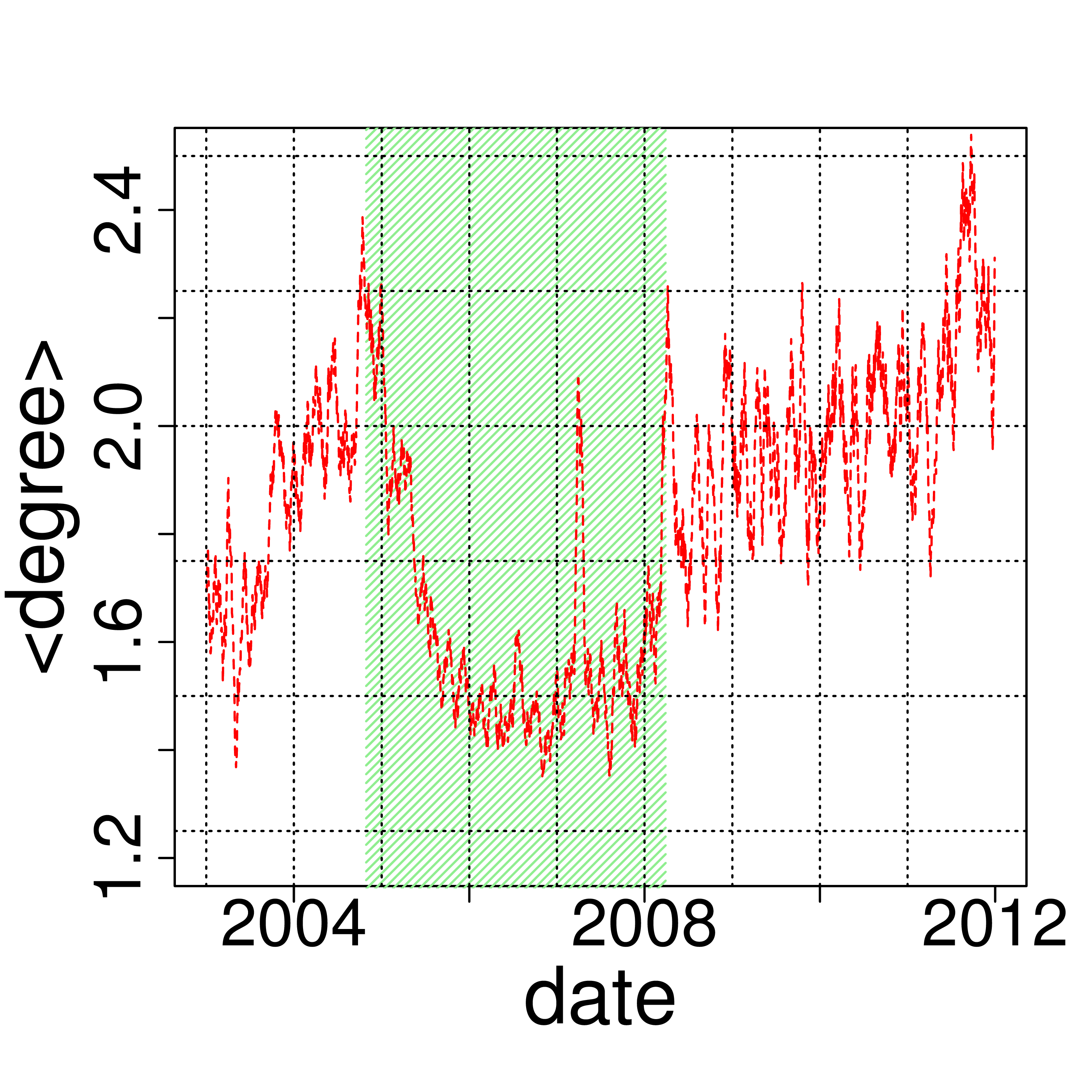}
}
\subfigure[algebraic connectivity\label{fig:structure:algcon}]{
  \includegraphics[width=0.3\textwidth]{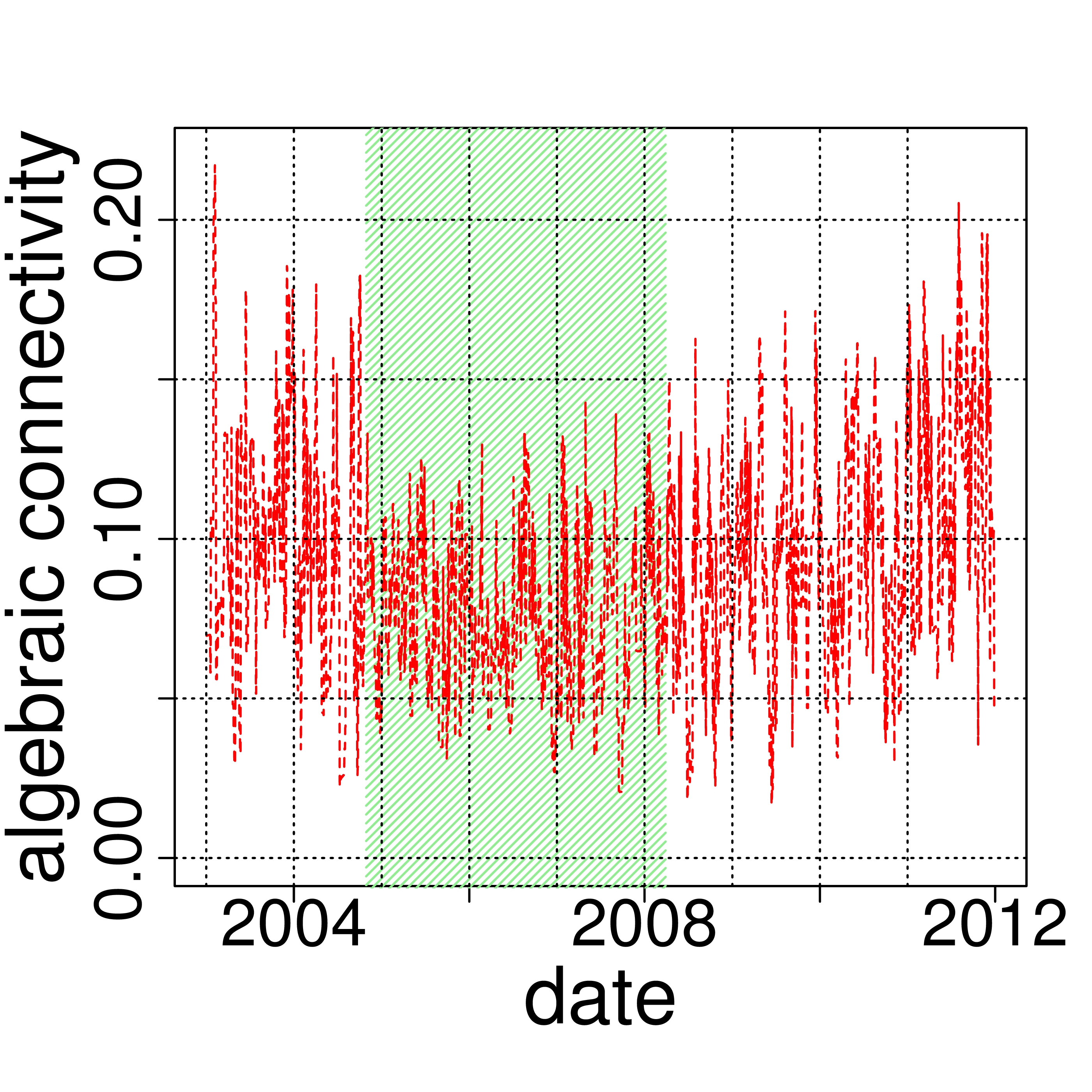}
}
\subfigure[algebraic con. (without \emph{Alice})\label{fig:structure_wo_alice:algconn}]{
  \includegraphics[width=0.3\textwidth]{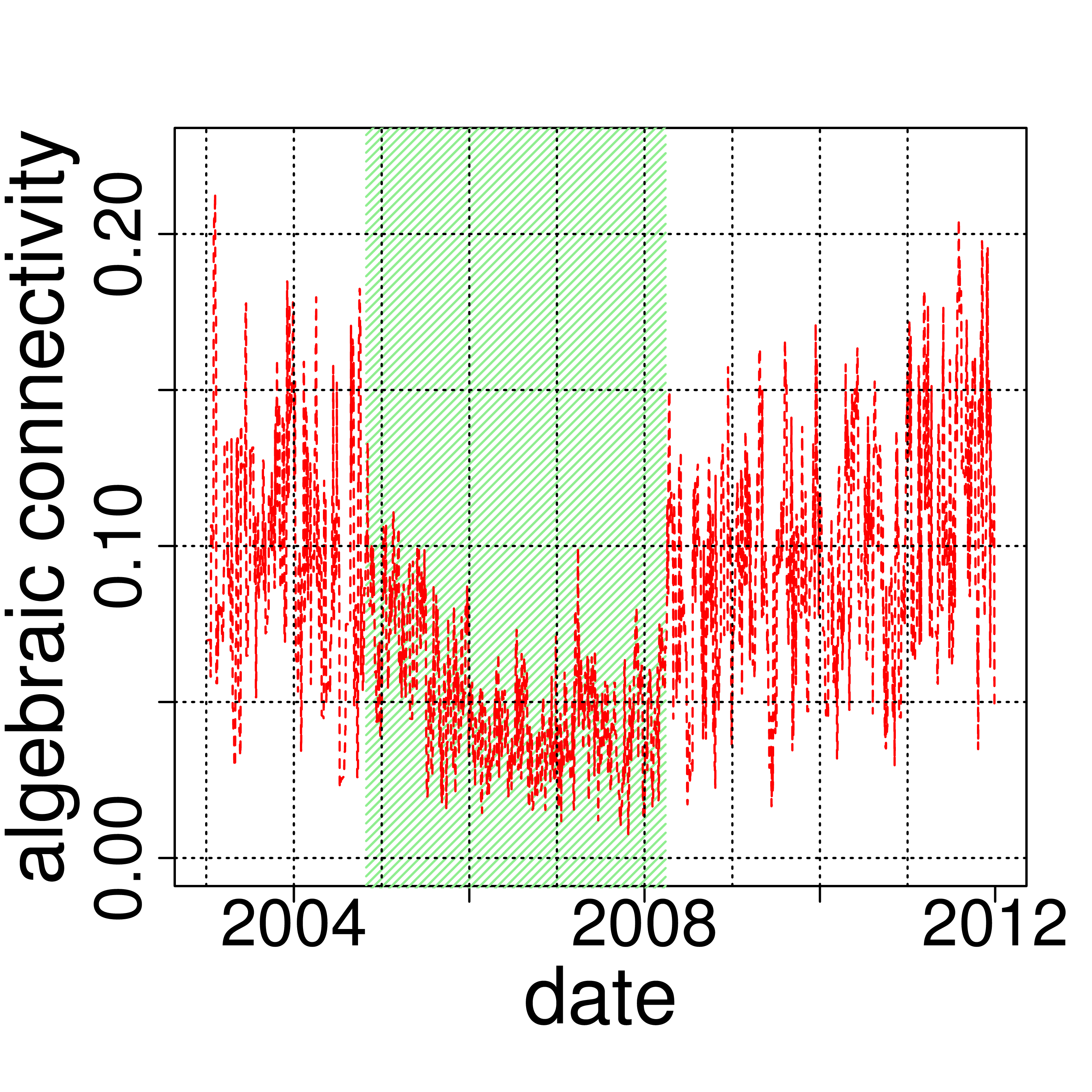}
}
\\
\subfigure[clustering coefficient\label{fig:structure:cc}]{
  \includegraphics[width=0.3\textwidth]{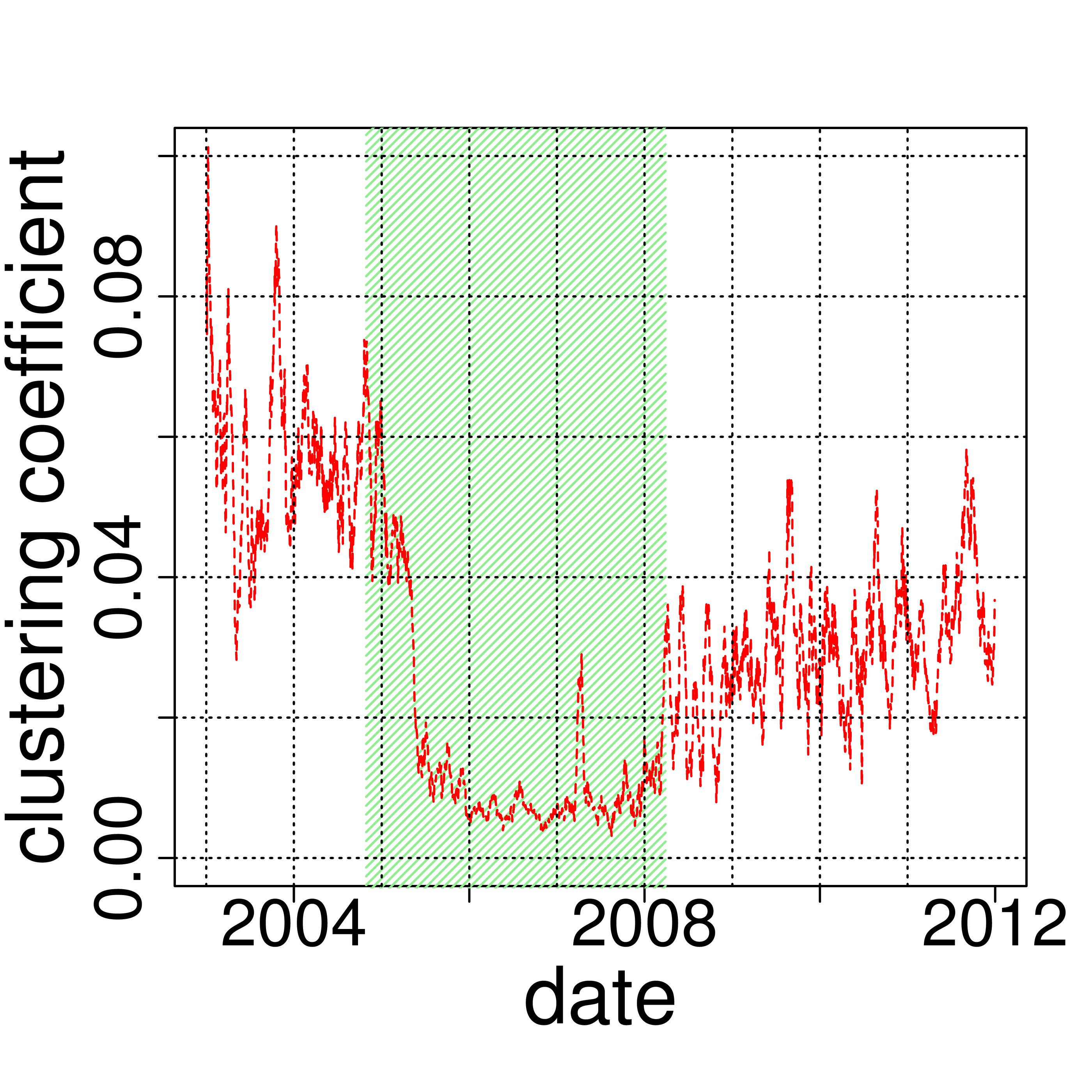}
}
\subfigure[assortativity\label{fig:structure:ass}]{
  \includegraphics[width=0.3\textwidth]{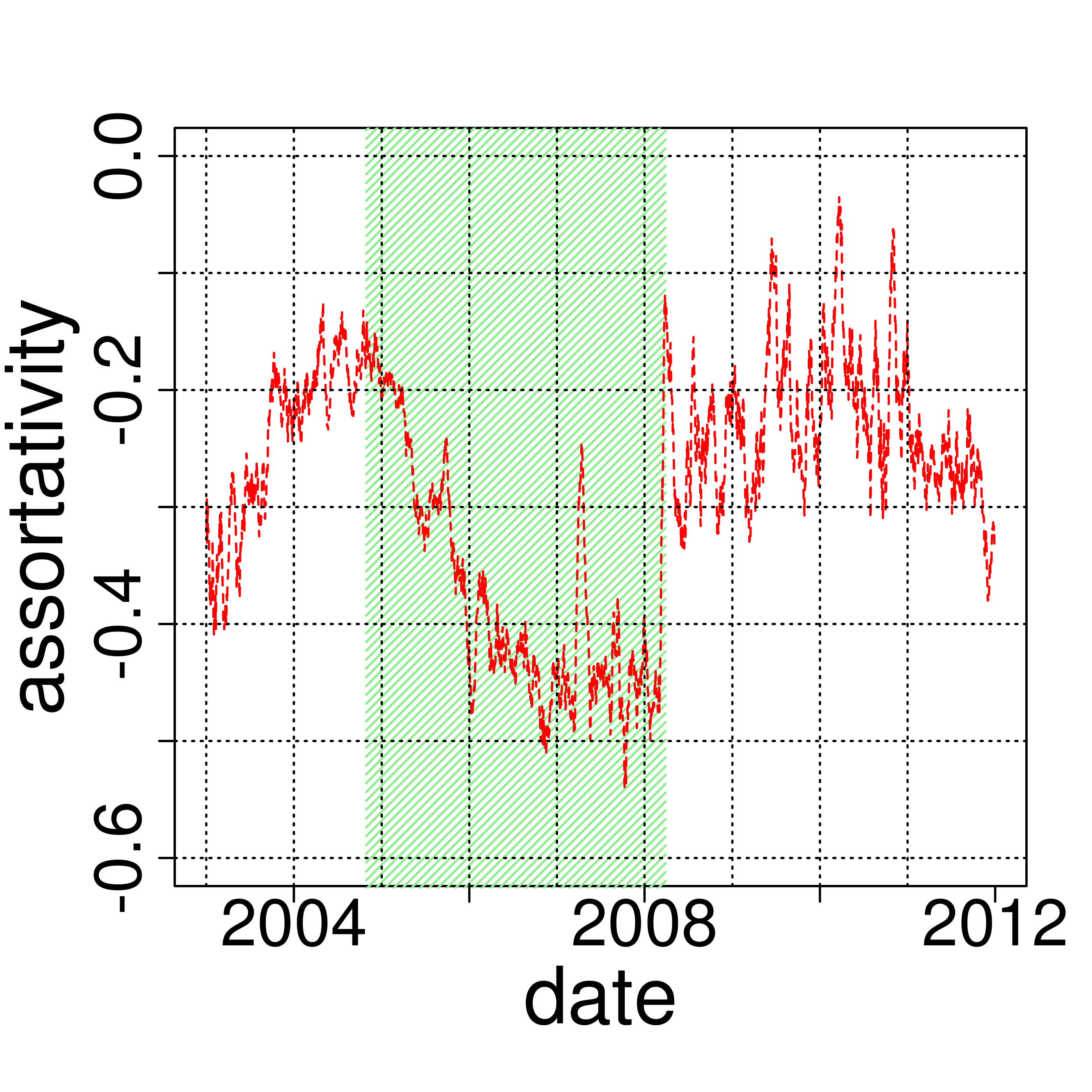}
} 
\caption{Dynamics of size and cohesion of the \textsc{Gentoo} bug handling community. Period $P2$ during which the central contributor \emph{Alice} was active is highlighted in green.\label{fig:structure}}
\end{figure}

We first focus on the size of the largest connected component (LCC) of the respective monthly collaboration networks.
The relative size of the LCC (i.e. the fraction of all nodes belonging to the LCC) is shown in Figure \ref{fig:structure:lcc_fraction}.
Since it captures how many of the contributors were disconnected from the rest of the community, this measure can be seen as a proxy for the \emph{cohesion} of the community.
In Figure \ref{fig:structure:lcc_fraction}, period $P2$ is highlighted in green.
As one can see, there is no significant difference between the periods $P1$ and $P2$ in terms of the relative size of the LCC; it rather remains stable around a value of $75 \%$.
However, a remarkable dynamics can be seen in period $P3$ after \emph{Alice} had left the community:
After a small drop, one observes a steady increase in the relative size of the LCC starting around the end of 2008.
The relative size of the LCC eventually reaches $95 \%$ around the end of 2011.

Another \emph{cohesion}-related measure is the average node degree in the monthly collaboration networks, i.e. the average number of different community members, a contributor was collaborating with during one month.
In Figure \ref{fig:structure:md}, one observes a fast decrease of this measure during period $P2$, when the central contributor \emph{Alice} was active.
Remarkably, it was \emph{increasing} during the periods $P1$ and $P3$, when \emph{Alice} was not active.
  
Apart from the relative \emph{size} of the LCC, a further interesting question is how efficient and robust the collaboration structures are \emph{within} the largest connected component.
For this, we compute the \emph{algebraic connectivity} of the LCC, a measure from spectral graph theory that captures how \emph{well-connected} a topology is.
As argued in section \ref{sec:methods}, networks with larger algebraic connectivity a) facilitate information flow and synchronization processes and b) are more robust against the loss of nodes and links.
The dynamics of algebraic connectivity is shown in Figure \ref{fig:structure:algcon}.
Comparing period $P2$ to $P1$ and $P3$, one observes that the presence of \emph{Alice} decreased both the variance and the mean of the algebraic connectivity.
A straight-forward interpretation of this finding is that - as \emph{Alice} was involved in many of the collaborations - the collaboration network's robustness decreased.
Furthermore, as most collaborations were mediated through her, the potential of congestion in this particular node increased, thus effectively decreasing communication efficiency of the topology.

Another interesting question from the perspective of social organization is to what degree two contributors that collaborated with a third contributor also collaborated with each other.
This is captured by the clustering coefficient of a network, whose dynamics is shown in Figure \ref{fig:structure:cc}.
The dramatic decrease of the clustering coefficient during period $P2$ and the gradual increase in period $P3$ highlights the mediator role played by \emph{Alice}.
As \emph{Alice} was involved in most of the collaborations, direct connections between users collaborating with her seemingly became unnecessary.
Another signature of the community's tendency to preferentially collaborate with the most central collaborator can be seen in \ref{fig:structure:ass}.
As described in section \ref{sec:methods}, the assortativity captures the preference of contributors to collaborate with other contributors that are more or less important than themselves.
A significant decrease of assortavitity from about $-0.15$ to $-0.45$ can be seen in period $P2$ when \emph{Alice} was active.
This substantiates the assumption that most community members primarily collaborated with the most central collaborator while collaborations between contributors with similar importance decreased.

A particular concern one may have in the analysis presented above is that it is unclear to what extent it is the presence of \emph{Alice} that affects the dynamics of network measures.
One may suspect that it is the mere number of collaborations involving her that increasingly dominate the community, while the existing collaboration structures are left more or less untouched.
In order to avoid this pitfall, we additionally ran our analysis on all monthly collaboration networks, however removing \emph{Alice} as well as all interactions in which she was involved.
We then computed the relative size of the LCC and algebraic connectivity to the residual networks.
Compared to Figures \ref{fig:structure:lcc_fraction} and \ref{fig:structure:algcon}, a clear difference can only show up during period $P2$, if \emph{Alice}'s presence indeed impacted the residual collaboration structures.
The plots of the relative size of the LCC (Figure \ref{fig:structure_wo_alice:lcc_fraction}) and algebraic connectivity (Figure \ref{fig:structure_wo_alice:algconn}) of the residual collaboration networks highlight that the activity of \emph{Alice} during period $P2$ significantly changed the organization of the community.
We particularly observe that - for the residual network - the fraction of users connected to the LCC dropped significantly from about $75 \%$ to about $30 \%$ over a period of two years.
Furthermore, algebraic connectivity of the residual network experienced a significant drop, thus highlighting that during \emph{Alice}'s presence the residual collaboration topology became less well-connected.

\begin{figure}[!ht]
\centering
\subfigure[October 2004\label{fig:nets:2004_10}]{
  \includegraphics[width=0.35\textwidth]{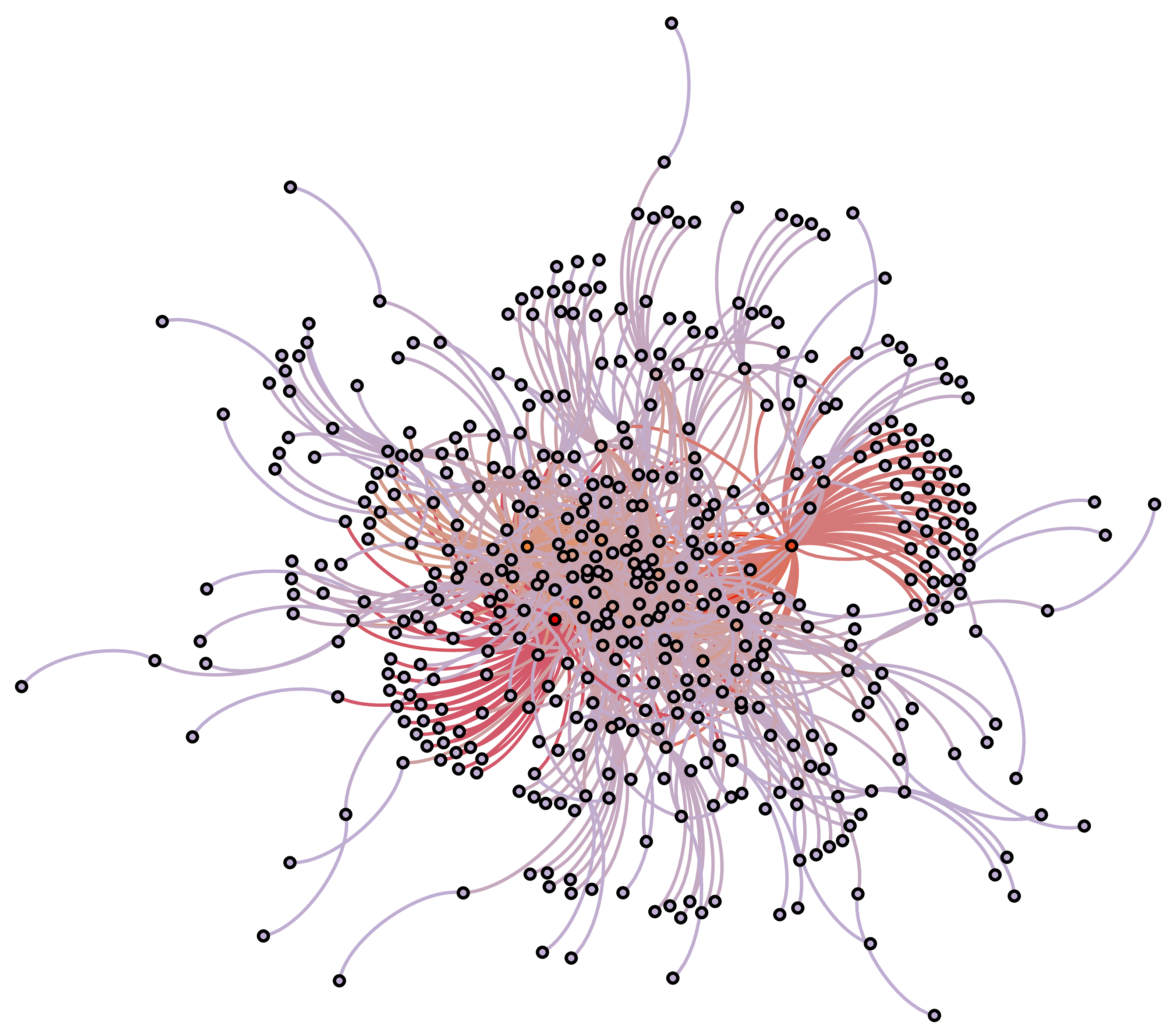}
}
\subfigure[July 2006\label{fig:nets:2006_07}]{
  \includegraphics[width=0.35\textwidth]{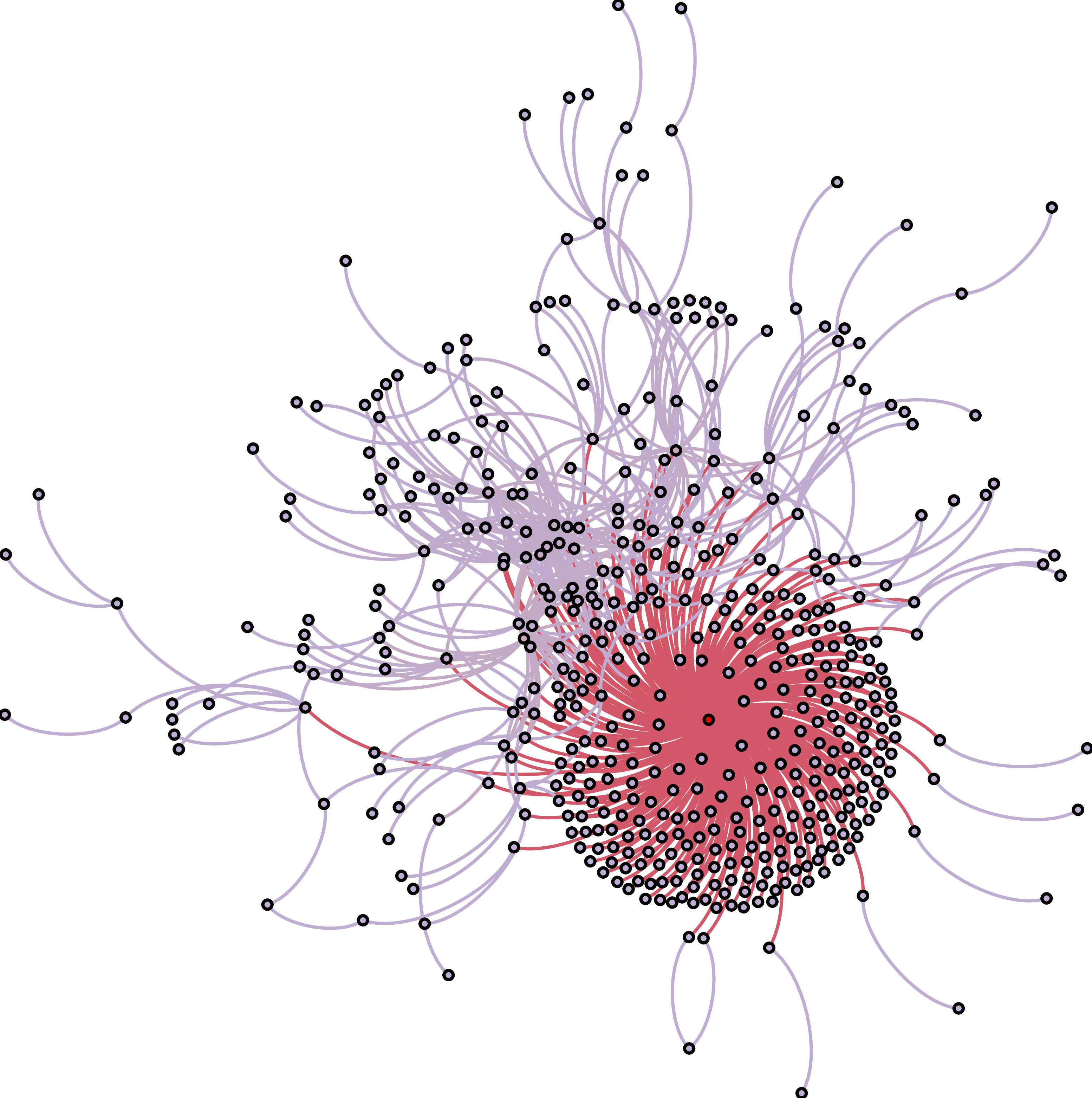}
}
\\
\subfigure[May 2008\label{fig:nets:2008_05}]{
  \includegraphics[width=0.35\textwidth]{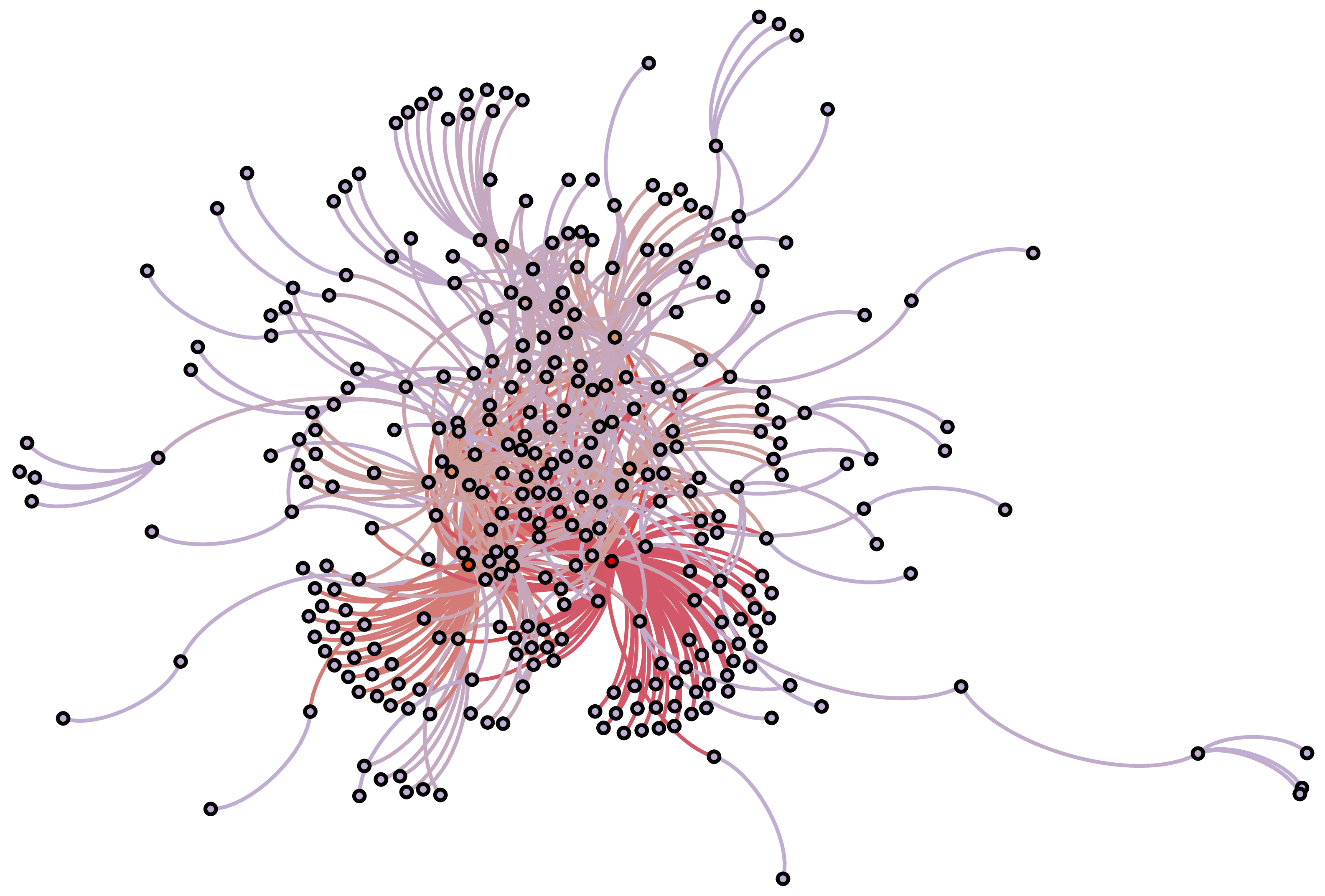}
}
\subfigure[July 2006 (without \emph{Alice})\label{fig:nets:2006_07_wo_alice}]{
  \includegraphics[width=0.35\textwidth]{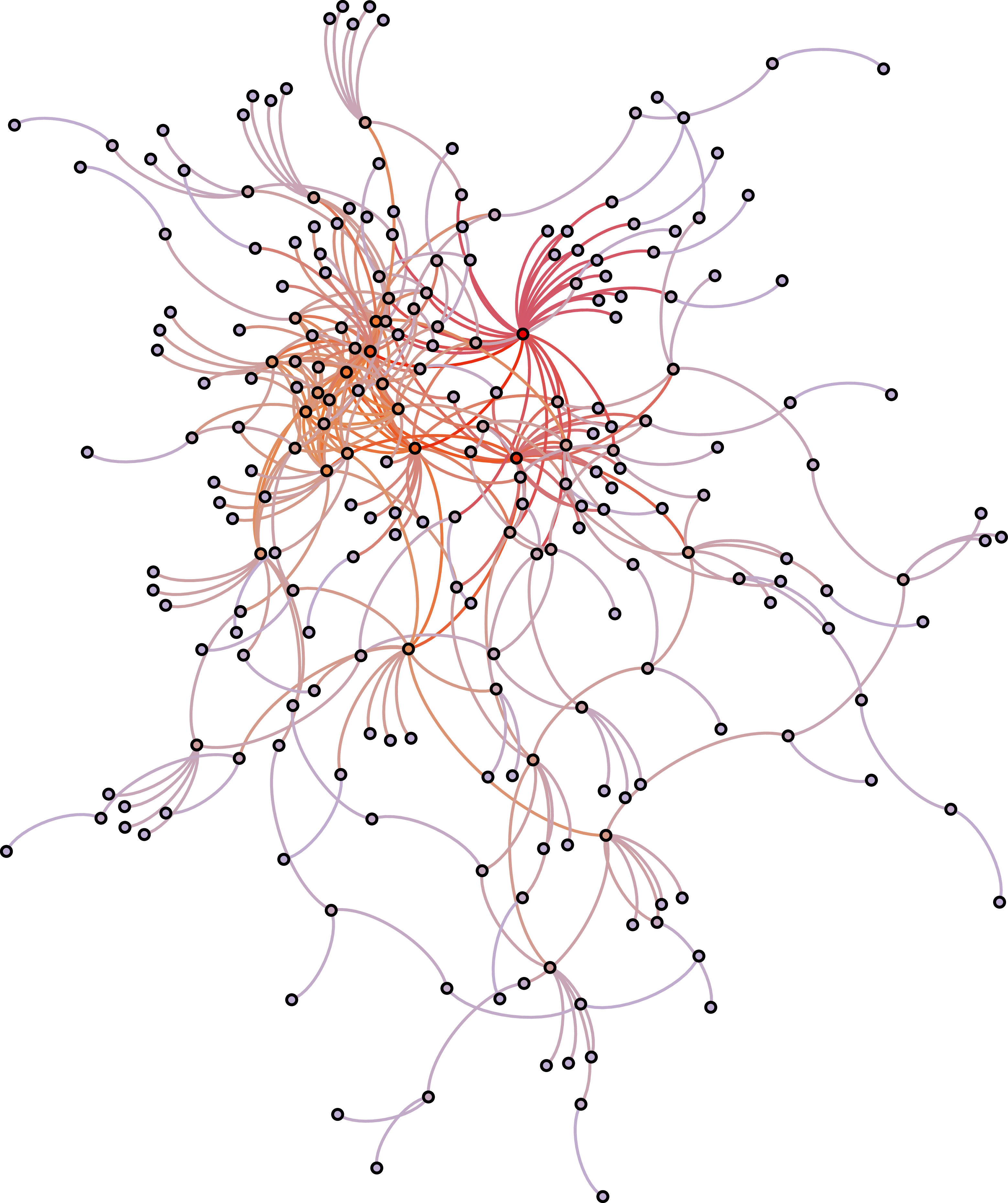}
}   
\caption{Illustration of representative monthly collaboration networks\label{fig:nets}}
\end{figure}

To visually illustrate the quantitative findings about the evolution of collaboration structures provided above, in Figure \ref{fig:nets} we additionally show four representative examples for the monthly collaboration networks during the periods $P1$ (Figure \ref{fig:nets:2004_10}), $P2$ (Figure \ref{fig:nets:2006_07}) and $P3$ (Figure \ref{fig:nets:2008_05}).
In addition, \ref{fig:nets:2006_07_wo_alice} depicts an example for a residual network constructed by removing all interactions of \emph{Alice} from the network depicted in Figure \ref{fig:nets:2006_07}.

From our quantitative study of the evolution of collaboration structures in the \emph{Gentoo} community, we can draw the following observation:

{\bf Observation:} {\em During the presence of the central contributor \emph{Alice}, cohesion in the \textsc{Gentoo} bug handling community decreased.}

\subsection{Centralization}
\label{sec:centralization}
 
A particularly important mechanism that could explain the loss of cohesion in the community is an increasing centralization of communication.
In this section we analyze the changes in centralization in the \textsc{Gentoo} community.
We not only study centralization from a network perspective, i.e. the increase of the \emph{topological} centrality of \emph{one} particular node.
We also consider the effects on the number of contributors that were involved in the bug handling process in terms of assigning bug reports or forwarding information.

\begin{figure}[!ht]
\centering
\subfigure[closeness centralization\label{fig:central:close}]{
  \includegraphics[width=0.3\textwidth]{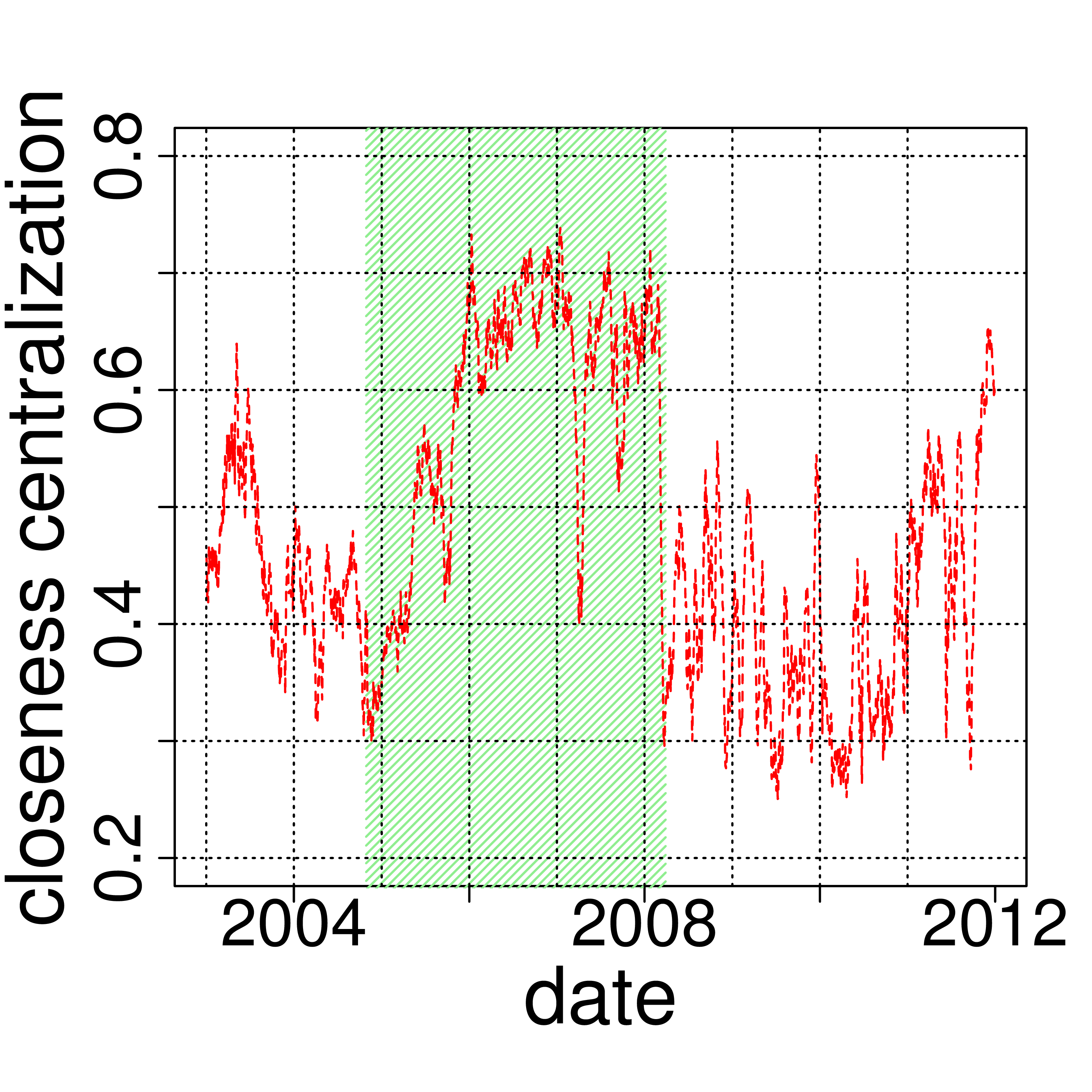}
}
\subfigure[contributors assigning bugs\label{fig:central:assigning}]{
  \includegraphics[width=0.3\textwidth]{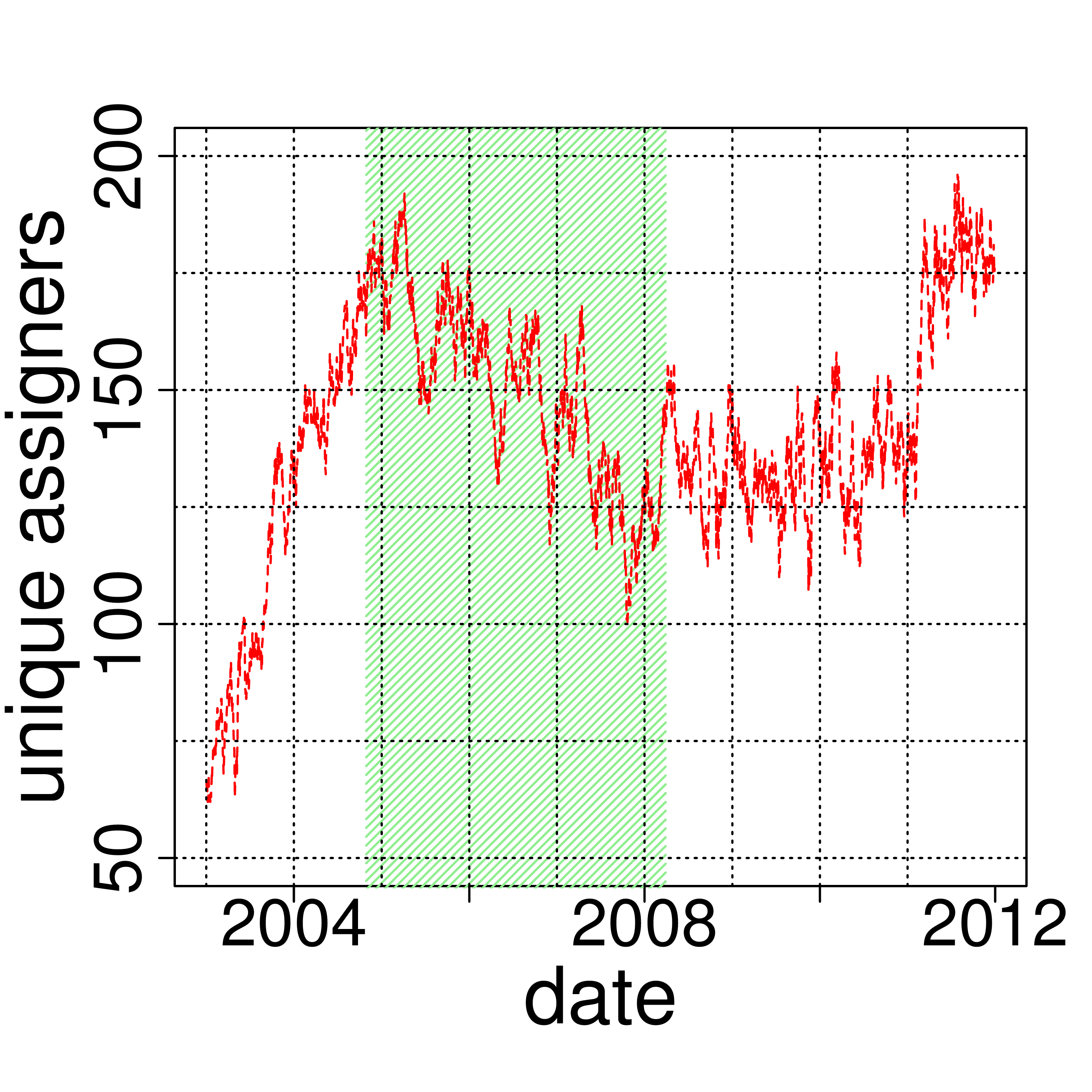}
}
\caption{Centralization in the \textsc{Gentoo} community. Period $P2$ during which the central contributor \emph{Alice} was active is highlighted in green.\label{fig:central}}
\end{figure}

We first analyze the degree of centralization in the \textsc{Gentoo} community from the perspective of \emph{closeness centralization}.
As argued in section \ref{sec:methods}, this measure captures to what extent the roles of contributors differ in terms of having short paths to all other community members.
The dynamics of closeness centralization shown in Figure \ref{fig:central:close} exhibits a decreasing tendency during the period $P1$.
A comparison to the dynamics of community size during $P1$ (see Figure \ref{fig:structure:size}) highlights that the growth of the community coincided with a decrease in centralization, which is in line with the findings of \cite{Crowston2005}.
However, the decrease in closeness centralization in period $P1$ was followed by a significant increase during period $P2$ when \emph{Alice} became active.
From the start of period $P2$ in October 2004 until the end in March 2008 closeness centralization increased from about $0.3$ to $0.7$.
When \emph{Alice} left the community, closeness centralization experienced a sudden drop, fluctuating around a value of $0.4$ during the period $P3$.

The finding that during period $P2$ the collaboration structures became more centralized is complemented by Figure \ref{fig:central:assigning}, which shows the number of different contributors assigning a bug report to another contributor within a given $30$ day period.
This number is of particular interest, as it captures how many contributors were actually involved in the bug triaging process by \emph{assigning} work to others.
Again mirroring the increasing size of the community, in period $P1$ one observes an increase in the number different contributors assigning bug reports.
At the end of period $P1$ in October 2004, about $170$ different contributors were assigning bug reports.
This increase is followed by a \emph{decrease} during period $P2$, again coinciding with the activity of \emph{Alice}.
This development was only stopped in March 2008, after \emph{Alice} had left the project.
After a sudden increase at the beginning of $P3$, the number of different contributors assigning bug reports remained rather stable until 2011, when it experienced another increase.

From the above analysis, we draw the following observation:

{\bf Observation:} {\em In the period where \emph{Alice} was active, centralization in the \textsc{Gentoo} community increased steadily.}

\subsection{Bug Handling Performance}
\label{sec:performance}

\begin{figure}[!ht]
\centering
\subfigure[daily activity\label{fig:perf:counts}]{
  \includegraphics[width=0.4\textwidth]{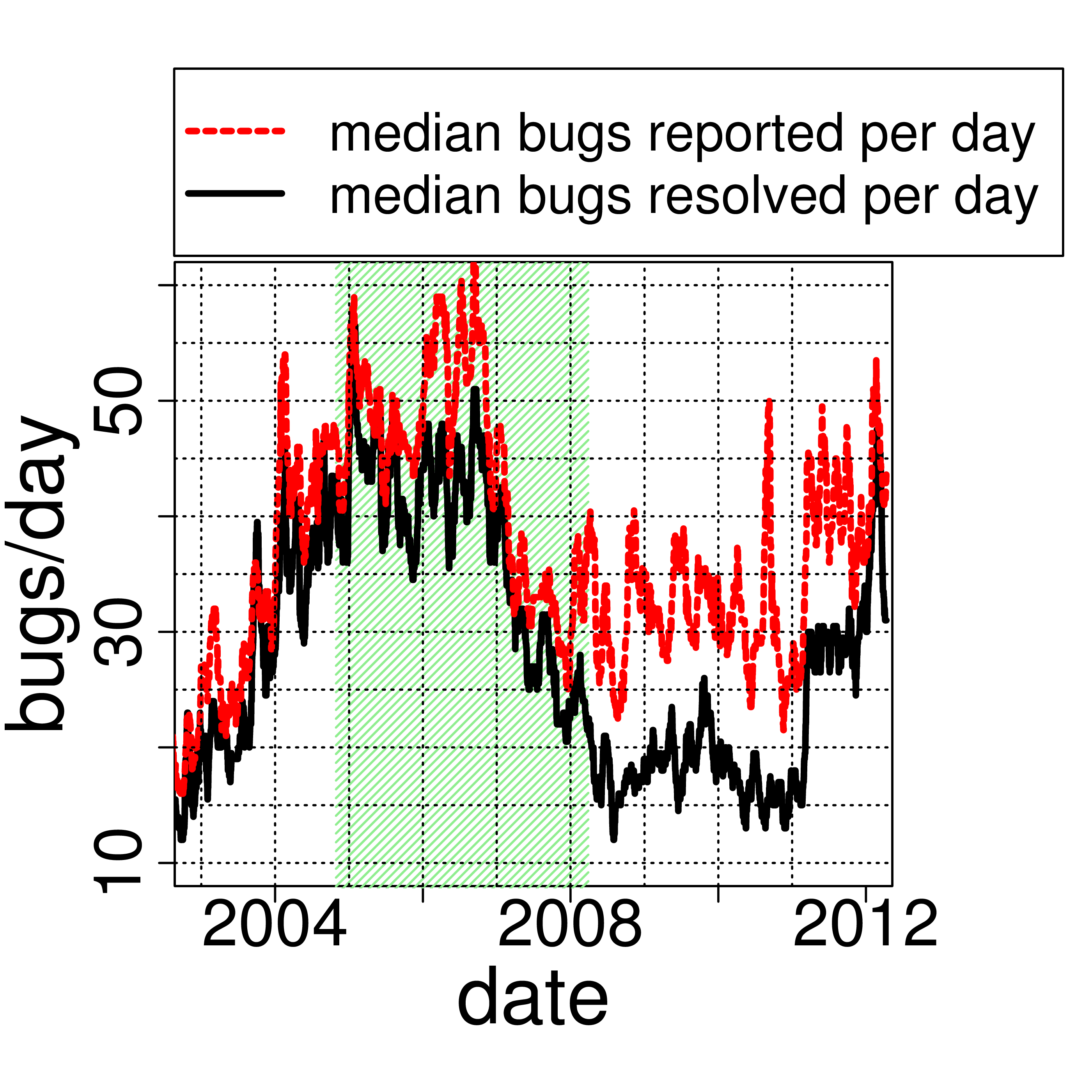}
}
\subfigure[time to first reply and resolution\label{fig:perf:resolved}]{
  \includegraphics[width=0.4\textwidth]{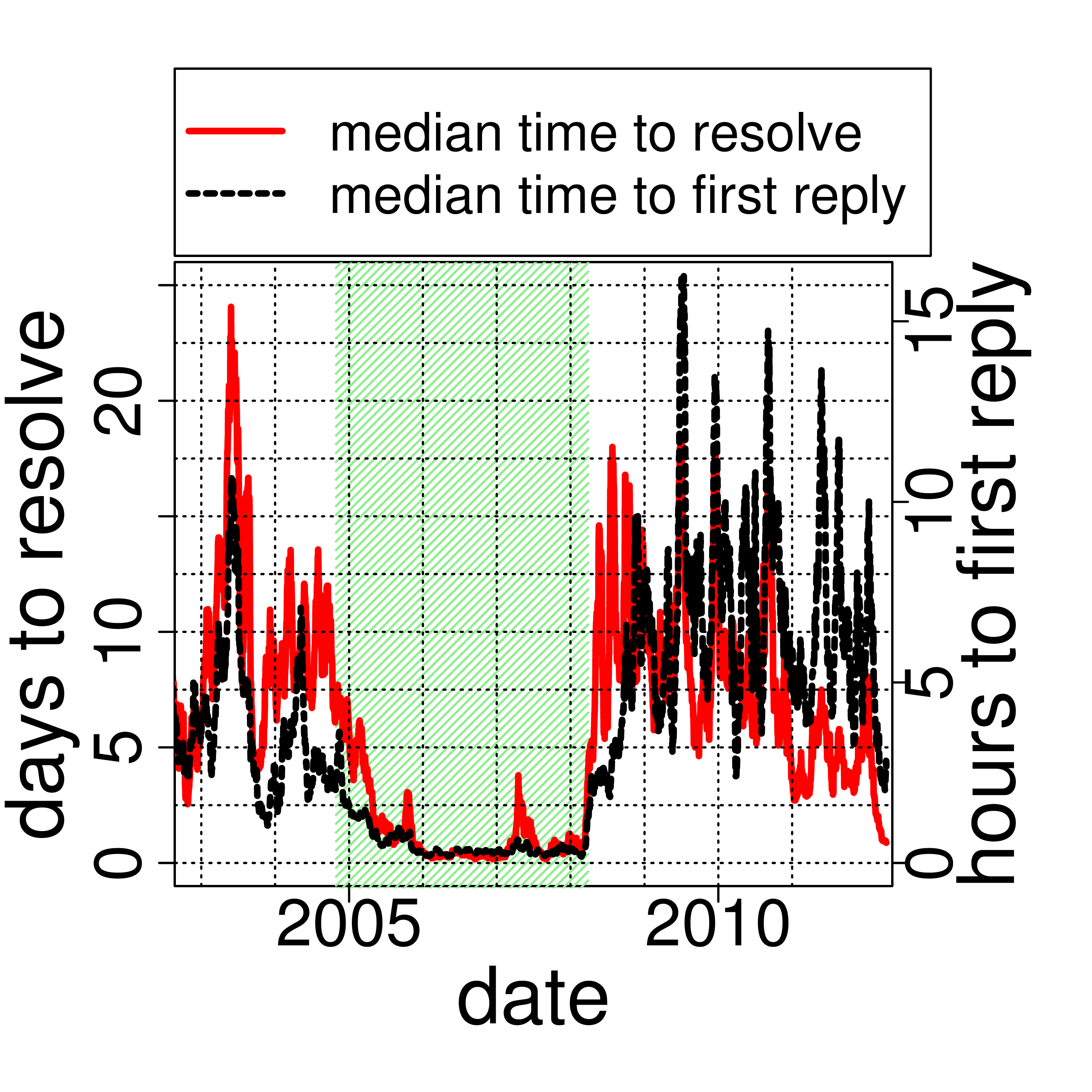}
}            
\caption{Bug handling performance in the \textsc{Gentoo} community. Period $P2$ during which the central contributor \emph{Alice} was active is highlighted in green.\label{fig:perf}}  
\end{figure}

Apart from studying the evolution of collaboration structures, our data set further allows to study the \emph{bug handling} performance of the \textsc{Gentoo} community.
As the simplest proxy for performance, we measure the rate at which bugs were reported and resolved.
We further study the responsiveness of the community in terms of the \emph{median time to resolve a bug}, i.e. the median time elapsed from the submission of a bug report to the point where it was finally resolved.
Similarly, we measure the \emph{median time to the first response} in terms of any update to the submitted bug report, like e.g. the bug being forwarded or assigned, commented on, or its status being changed to reproduced.
Figure \ref{fig:perf:counts} shows the dynamics of the median number of bugs that were reported and resolved per day.
During period $P1$ one observes a continuous increase both in the number of reported and resolved bugs which coincides with the growth of the \textsc{Gentoo} community shown in \ref{fig:structure:size}.
During period $P2$, both the number of reported and resolved bugs decreased, which can again be understood based on the decrease in the number of active contributors shown in Figure \ref{fig:structure:size}.
In both periods $P1$ and $P2$, the rate of reporting and resolving bugs closely match each other, thus indicating that - on average - the number of bugs resolved per day matched the number of newly reported bugs.
This lastingly changed after \emph{Alice} had left the project.
In period $P3$ one can observe an increasing discrepancy between the rate at which bugs were reported and resolved, hence indicating a growing number of unresolved, pending bug reports.
Furthermore, a remarkable increase in both the number of reported and resolved bug reports can be seen around March 2011, although the discrepancy between both remains.
This coincides with an increase in the number of active community members (see Figure \ref{fig:structure:size}).
One possible explanation is that it coincides with the \textsc{Gentoo} community having a regular \emph{LiveDVD} release.
As it lowers the threshold of using the \emph{Gentoo Linux} distribution, this can explain an increasing number of contributors submitting bug reports, as well as the increase in the number of different contributors assigning bug reports shown in Figure \ref{fig:central:assigning}.

Apart from the mere number of reported and resolved bugs, an important measure of performance of bug handling communities is the time they take to provide a first response as well as a resolution for a reported bug.
This \emph{responsiveness} is of particular importance, as potential users frequently use this as an indicator when making an informed decision about which software to adopt.
Figure \ref{fig:perf:resolved} shows the median time resolve and respond to a newly reported bug in days and hours respectively.
Both numbers show a remarkable dynamics which coincides with the activity of the central contributor \emph{Alice}.
During period $P2$, the median time to resolve and respond to a newly submitted bug report was more than one order of magnitude smaller than in the periods $P1$ and $P3$.

From our analysis of bug handling performance, we thus draw the following observation:

{\bf Observation:} {\em During the presence of the central contributor \emph{Alice}, the bug handling performance of the \textsc{Gentoo} community increased significantly, while her retirement had a lasting negative impact.}

\subsection{Discussion}
\label{sec:discussion}

We close this section by combining our quantitative results with personal insights shared by three long-term contributors: \emph{Alice}, \emph{Bob} and \emph{Chris}.
By this, we substantiate our interpretation of \emph{Alice}'s role during period $P2$ and the consequences of her presence for the cohesion and performance of the \textsc{Gentoo} community.

In sections \ref{sec:cohesion}, \ref{sec:centralization}, and \ref{sec:performance}, we point out that during period $P2$ the community experienced a significant loss of cohesion, as well as an increase of centralization and performance.
We further argued that during period $P2$, most of the collaboration was mediated by a small subset of contributors, \emph{Alice} herself being at the core of this group.
Indeed, in response to our questionnaire, \emph{Alice} describes that she ``was practically the only person involved in bug wrangling''. \emph{Bob} confirms that \emph{Alice} ``had been doing our \emph{bug wrangling} more or less alone for a few years''.
\emph{Alice} complements this picture by saying that the ``workload at that time - [if I recall correctly] - was about 4 hours a day, probably more in case I did not have time to do the bug wrangling for a day or two''.
As a consequence of this centralization of bug handling tasks, during period $P2$ our analysis shows a significant increase in performance, measured in terms of response time and bug resolution rate.
This finding is confirmed by the community and \emph{Bob} attributes it to the fact that ``having a single person on the task greatly helps in finding duplicate bug reports''. 
Furthermore, he argues that having ``more [contributors] would water down the quality''.

A further observation of our study is that the cohesion of the community (measured e.g. in terms of mean degree, clustering coefficient or algebraic connectivity) decreased significantly during the presence of \emph{Alice}.
This is an interesting observation as it highlights secondary effects of the presence of a central contributor on the evolution of collaboration structures within the remaining community.
Although it is necessarily difficult to make any substantiated claims about causality, one may conjecture that it is the mere presence and dedication of a central contributor that drives this loss of cohesion.
\emph{Bob} indirectly confirms this by arguing that apparently ``our bug tracker's users had come to rely on a single person to ``assist'' them in finding and fixing bugs''.

For the community, the retirement of \emph{Alice} was perceived as an unexpected event. 
According to \emph{Bob}, in 2008 \emph{Alice} ``suddenly left the project''. He further confirms that she ``stopped unexpectedly''.
Clearly, one of the most interesting questions that cannot be answered by a quantitative study alone is why \emph{Alice} decided to leave the community.
She answered our question for the underlying reasons as follows: ``I would mostly attribute that to a serious loss of motivation caused by disruptive social environment in the project as a whole''.
Moreover, she highlights her dissatisfaction with ``more and more time being spent on bureaucracy, "paperwork", and creating of useless structures within the project''.
On the contrary, \emph{Chris} - another prominent contributor - remarks that ``some people find formalization to be an unnecessary bureaucratic barrier, but when you get to be as big as Gentoo, it's pretty much inevitable''.

Independently of the reasons for \emph{Alice}'s retirement, the risk of relying too much on a central contributor became obvious in a remarkable event during period $P2$, when \emph{Alice} was still active.
In early 2007, according to her own account, \emph{Alice} was ``repeatedly subject to [...] disciplinary proceedings and [she] was suspended from the project for a couple of weeks'' due to a verbal conflict with another contributor.
Around this time, a sudden and short increase in the response time (see Figure \ref{fig:perf:resolved}) as well as a decrease in closeness centralization (see Figure \ref{fig:central:close} can be observed, thus serving as an early warning sign of the problems to come when \emph{Alice} would leave.

Despite this early indicator, it was only after \emph{Alice} had left that the community took measures to reorganize the community.
In particular, \emph{Bob} initiated the \textsc{Bug Wranglers} project, which a) called for more contributors in bug handling and b) established formal procedures regarding the tasks and goals of bug triaging\footnote{See the website of the \textsc{Bug Wranglers} project available online at \url{http://www.gentoo.org/proj/en/qa/bug-wranglers/index.xml}}.
In response to our questions, \emph{Bob} describes the project as a success arguing that ``the targets that relate to the content of bug reports are now usually met when serious bug wranglers review them''.
However, despite this initiative, our finding of a lasting negative impact on bug handling performance after the resignation of \emph{Alice} is confirmed by \emph{Bob}, saying that the ``goal of responding to bugs within a day is still something to work on''.

\subsection{Threats to Validity}
\label{sec:threats}
We now discuss limitations of our analysis and highlight possible threats to validity.
Since our paper is a case study focused on the \textsc{Gentoo} community, we cannot make any claims about the general applicability of our results. 
Even though our study as well as the feedback by the community provide some interesting hints, we would further like to emphasize that we cannot make conclusive statements regarding the causal relation between increasing centralization, performance and cohesion. 
In particular, we cannot rule out external reasons driving \emph{both} the increase of centralization and the loss of cohesion in the community.
Despite this disadvantage, we argue that our case study is interesting by itself, being a valuable addition to the literature on benefits and risks of centralization in collaboration topologies.
In order to validate our findings, we thus call for similar studies on OSS communities and other collaborative software engineering projects.

Another possible concern is the choice of our network construction procedure as well as the choice of length of the sliding window in our dynamic analysis. 
In order to only extract \emph{meaningful} collaboration events and facilitated by the size of our data set, we only considered \emph{cc} and \emph{assign} collaborations. 
Nevertheless, it is clear that taking into account further relations, like e.g. comments, could possibly augment our perspective of collaboration topologies. 
At the same time, we argue that - even though we have explored different sizes for the sliding window - we did not see any qualitative change of our results. 
Eventually, we decided to include the results of a $30$ day window size, since this period is long enough to include collaborations of more occasional collaborators.
At the same time, a one month period is short enough to not aggregate collaborations occurring far apart in time. 
As such, our methodology of performing a \emph{dynamic network analysis} can be seen as a strength compared to the simpler approach of considered a single time-aggregated network. 
                    
\section{Conclusion}
\label{sec:conclusion}

The main contributions of our paper are the following:
\begin{itemize}
\item We study the dynamics of social organization and performance in the bug handling community of \textsc{Gentoo}.
\item We find a period in which the activity of a single contributor resulted in a significant increase of centralization and performance.
\item Our analysis further shows that the period when the central contributor was active coincided with a significant decrease of cohesion.
\item We further find that the loss of the central contributor had a lasting negative impact on the bug handling performance of the community. 
\end{itemize} 

To the best of our knowledge, our paper is the first to quantitatively study how the rise and fall of a central contributor impact the social organization and performance of an OSS community.  
Even though the general statements that can be drawn from a case study are necessarily limited, we argue that our work highlights interesting directions for future research.
We would like to emphasize that the quantitative measures used in our study allow to clearly identify shifts in the social organization that are confirmed by insights by actual contributors.
As such, we argue that these measures can potentially be used in monitoring tools suitable to augment the social awareness of community managers.
   
\section*{Acknowledgment}

This work was supported by the SNF through grant CR12I1\_125298 and by the European Commission's Seventh Framework Programme FP7-ICT-2008-3 under grant agreement No. 231323 (CYBEREMOTIONS). 
We acknowledge the contribution of Emre Sarig\"ol to data collection and processing and would like to thank the \textsc{Gentoo} community members \emph{Alice}, \emph{Bob} and \emph{Chris} for sharing personal insights with us.

\bibliographystyle{acm}
\bibliography{paper}
    
\end{document}